\documentclass[aps,floatfix,prd,showpacs]{revtex4}
\usepackage{graphicx}
\usepackage{dcolumn}
\usepackage{bm}

\begin{document}

\def\dq{\frac{d^4q}{(2\pi)^4}\,}
\def\dqE{\frac{d^4q_E}{(2\pi)^4}\,}
\def\be{\begin{equation}}
\def\ee{\end{equation}}

\def\capstyle{\rm}

\def\A{\capstyle A}
\def\B{\capstyle B}
\def\C{\capstyle C}
\def\D{\capstyle D}
\def\G{\capstyle G}
\def\H{\capstyle H}
\def\I{\capstyle I}
\def\J{\capstyle J}
\def\K{\capstyle K}
\def\L{\capstyle L}
\def\P{\capstyle P}
\def\X{\capstyle X}
\def\Y{\capstyle Y}
\def\Z{\capstyle Z}

\def\NN{\capstyle NN}
\def\JP{\J^{\P}}
\def\JPC{\J^{\P\C}}
\def\JPG{\J^{\P\G}}

\def\1S{\capstyle 1S}
\def\2S{\capstyle 2S}
\def\3P0{$^3$P$_0$}

\title{Molecular Interpretation of the Supercharmonium State $\Z(4475)$}

\author{T. Barnes}
\affiliation{
U.S. Department of Energy,
Office of Science,
Office of Nuclear Physics,
Washington, DC 20585,
USA.}

\author{F. E. Close}
\affiliation{
Rudolf Peierls Centre for Theoretical Physics, University of Oxford, 
Oxford, OX1 3NP, UK.}

\author{E. S. Swanson}
\affiliation{
Department of Physics and Astronomy,
University of Pittsburgh,
Pittsburgh, PA 15260,
USA.}

\date{\today}

\begin{abstract}
Charm meson molecule $(c\bar n)(n\bar c\, )$ assignments for the supercharmonium state $\Z(4475)$ 
are considered, both on general grounds and in the context of specific pion exchange models. 
Two possible charm molecule assignments for the $\JP = 1^+$ $\Z(4475)$ are considered here, 
involving respectively a radially and an orbitally excited charm meson. In each assignment, 
a lower-mass isosinglet charm molecule state is predicted. For both the $\Z(4475)$ and the $\Y(4260)$, 
measurement of the ratios of the branching fractions to ${\D}\bar{\D}$ plus $n$ pions is recommended 
as a test of the nature of these states.
\end{abstract}

\maketitle

\section{Introduction}

The $\Z(4475)$, a remarkable hadronic resonance with a mass and production properties reminiscent of 
charmonia, but which carries electric charge, is clear evidence for a state that is qualitatively 
different from the known ``$c\bar c$ quark model" charmonium resonances \cite{4475}. 
The detection and recent confirmation of the $\Z(4475)$ and measurement of its $\JP = 1^+$ quantum numbers 
have excited considerable interest \cite{interest}. 

Since the $\Z(4475)$ is produced in a charmonium-rich B meson weak decay process, has a mass typical of 
excited charmonia, decays to a charmonium channel ($\psi' \pi$), and yet has net electric charge, 
there is strong evidence that this state is the product of charmonium $c\bar c$ and light-quark $d\bar u$ 
basis states. Unfortunately, four-quark states occupy a long, complicated, and rather 
controversial chapter in the history of hadron physics, and the terminology used in the recent $\Z(4475)$ 
literature has occasionally been confusing or ambiguous. 

In an attempt to rationalize the terminology, we propose the generic term ``supercharmonium" for an 
experimental resonance that appears to contain a $c\bar c$ pair, but has other properties that preclude 
a description in terms of only $c\bar c$ (idealized charmonium) basis states. Within this general class 
of state, the remaining questions involve the Fock space decomposition of the state, the nature of 
the non-$c\bar c$ components, the type of spatial wavefunctions involved, and the dynamical origin 
of the state. Here, in the interest of clarity, we will briefly discuss various possible subclasses 
of such ``superonia" (resonances beyond idealized $q\bar q$ quarkonia). 

When one considers the first Fock-space color-singlet quark model meson basis states beyond $q\bar q$, 
generically $q^2\bar q^2$, the associated terms ``four-quark state" and ``tetraquark" appear {\it prima facie} 
to simply specify that $q^2\bar q^2$ is the dominant Fock-space state present in the decomposition of 
a state. However, these terms have historically come to refer more exclusively to a specific type of 
$q^2\bar q^2$ state, in which the spatial wavefunctions of the system describe a closely overlapping 
tetraquark cluster. The classic examples of such states are provided by the MIT bag model \cite{jaffe}, 
in which all the quarks and antiquarks are distributed relatively uniformly through a common spherical 
region of extent $1/\Lambda_{QCD}$. Such states are known rather confusingly as ``baryonia" when composed of light $q=u,d,s$ quark 
flavors, because it was once thought that they would be made copiously in baryon-antibaryon annihilation, 
$q^3\bar q^3 \to qq\bar q\bar q$. 

A well-known argument for the nonexistence of such four-quark cluster states as physical resonances was provided 
by Coleman in 1979 \cite{1/n}, who used the $1/N_c$ expansion of QCD to show that four-quark color singlets 
tend to propagate as pairs of mesons. However, Weinberg has pointed out that subdominance of a four-quark 
component does not mean that it is irrelevant; one must rather examine the lifetime of the component 
to determine its viability \cite{weinberg}.  In this regard, Weinberg noted that the $f_0(980)$ member 
of the low-lying scalar nonet evades Coleman's argument because it lies near $\K\bar{\K}$ threshold, 
whereas the $f_0(500)$ could be an example of a four-quark state that exists despite its large width. 
Generally, one expects that observable tetraquark systems arise with small or normal widths when either 
(i) the only decay mode is to a pair of mesons near threshold, or 
(ii) when the fall-apart modes are kinematically forbidden, so that decays require $q\bar{q}$ creation 
or OZI violation.

Several qualitatively different possibilities for the spatial wavefunctions associated with (non-cluster) 
four-quark or tetraquark basis states have been suggested. Here we mention three such possibilities 
proposed in the literature: diquonium, hadrocharmonium, and two-meson molecules. Of course, not all 
experimental enhancements with exotic characteristics need be superonia. For example, there is good evidence that the $\Z_c(3900)$,  $\Z_b(10610)$, and $\Z_b(10650)$ are coupled-channel cusp effects, and not truly resonances \cite{Zcusp}.

In the extreme case of diquark--antidiquark \cite{diq} ``diquonium" systems $(qq)(\bar q \bar q)$ that 
are spatially separated by a large angular barrier, it was noted that the fall-apart effect might be 
so suppressed that four-quark resonances could be observed. However the orbital angular momentum 
$\L$ between the $(qq)$ and $(\bar q \bar q)$ subsystems required to suppress fall-apart was estimated 
to be sufficiently large to discourage further studies of this system.

Another possibility for a generalized charmonium state was discussed by Voloshin \cite{DV}. This work 
postulates an important chromoelectric dipole interaction between a charmonium core and light hadronic 
matter, thereby creating a supercharmonium state, which was referred to as ``hadrocharmonium." 

Finally, a widely discussed type of four-quark state, which includes $c\bar c$ in the dominant basis state, 
is a ``charm meson molecule." This is (naively) a $c\bar q$ and $q'\bar c$ charm meson pair that is 
relatively weakly bound by the residual strong force so that the wavefunction is close to that of 
two unperturbed mesons. ``Hadronic molecule" states of this general type are of course very well 
established, since they include nuclei and hypernuclei. Unfortunately, two-hadron interactions are 
generally rather poorly understood, so it is difficult to anticipate which two-meson systems should 
have sufficiently attractive interactions to form molecules \cite{isgurweinstein}. As general guidelines, 
since these forces are short ranged, S-wave meson pairs should be the most favorable possibilities, 
and typical nuclear binding energies of 1-10 MeV scale might be anticipated.  

The charm meson molecule candidate $\X(3872)$ is a well known example, as its mass is very close to the 
${\D}^0\bar {\D}^{*0}$ two-meson threshold, and it has the $\JP = 1^+$ quantum numbers expected for this 
meson pair in S-wave \cite{3772data}. The $\X(3872)$ is often regarded as a hadronic molecule bound 
by $\pi$-exchange forces. That such states should occur was predicted in Ref. \cite{torn}, and detailed 
dynamical models were first developed in Refs. \cite{cp, swanson}.

Another molecular candidate is the vector state $\Y(4260)$. This is produced in $e^+e^-$ annihilation, 
has a width of about 90 MeV, and has only been seen in the mode $\psi \pi\pi$ \cite{y4260}. The decay mode 
is superficially OZI violating, yet the width is canonically hadronic, and not approximately 1 MeV as 
might have been expected for a conventional OZI-violating channel. An energy of 4260 MeV is the first 
place where charmed meson pairs can be produced from $e^+e^-$ annihilation in S-wave, namely the 
$0^-1^+$ combination $\bar{\D} {\D}_1$ or the $1^-0^+$ combination $\bar{{\D}^*}{\D}_0$. A conservative 
interpretation \cite{fc4260} is that the $e^+e^-$ annihilation produces a $c\bar{c}$ pair, followed by 
the strong production $u\bar{u}$ (say) for the anticipated standard decay. The relative momentum being 
near zero, and the $\D$ being stable on the time scales of strong interactions, allows the initially 
produced tetraquark state to rearrange as $[c\bar{u}][\bar{c}u] \to [c\bar{c}][u\bar{u}]$, which permits 
the $\psi (\pi\pi)$ discovery mode. It is possible for $\pi$-exchange forces to play a role here too, 
through the off-diagonal coupling $\bar{\D} {\D}_1 \to \bar{{\D}^*} {\D}_0$; this could allow both 
$1^{--}$ and exotic $1^{-+}$ states to arise.

The anomalously low-lying charm-strange $0^+$ state at 2310 MeV is also intriguing. It has been suggested by Barnes, Close, and Lipkin that this is a $\D\bar{\K}$ analog of the $f_0$ and $a_0(980)$ $\K \bar {\K}$ molecule candidates\cite{bcl}. These authors also predicted the existence of a ${\D}^*\bar{\K}$ $1^+ $ molecular near 2460 MeV, which appears to have been found \cite{1+}.

A common feature of all the molecular candidates in the charmonium mass range is that a four-quark 
supercharmonium configuration occurs very near, or even below, the lowest (S-wave) threshold for production 
of meson pairs. Thus these states all evade Coleman's argument. This mechanism is not universal across 
all flavors because the meson pairs into which the superonia would preferentially decay are pseudoscalar 
mesons with masses that are lowered by chromomagnetic effects, which scale as $1/E_iE_j$ \cite{ss}. Here 
$i$ and $j$ refer to meson constituents and $E$ refers to their masses (in a constituent picture) or energies 
(in an MIT bag picture). Thus pseudoscalar mass shifts are reduced as the quarks become heavier. The mass of 
the superonium is less trivially dependent on these constituent ``masses", as it depends on both the 
various color substates and spin states of the quark and antiquark constituents. So while we understand 
why the coincidences between a superonium mass and the channel for two pseudoscalar mesons are not universal, 
it is less clear how to make sharp statements as to when and where such coincidences might happen, so that 
observable superonium states appear. The moral, however, is that these delicate balances arise as a result 
of interquark forces, which are not taken into account in the $1/N_c$ arguments of Coleman. These perturbations 
within the hadrons can significantly modify the Coleman arguments, so that observable tetraquarks can occur.

The challenge is to understand the dynamical origin of various four-quark superonia, with the goal of 
classifying families of states.  It is clear that there is no common mechanism responsible for these 
states. In the following we shall consider two possible charm meson molecule assignments for the 
$\Z(4475)$ in the context of a simple pion exchange model of the intermeson nuclear force.

If two hadrons $\A$ and $\B$ are linked by $\A \to \B \pi$, then hadronic pairs $\A\B$ or 
$\A\bar{\B}$ necessarily experience a pion exchange interaction. This force will be attractive in at least 
some channels. Long ago, the idea of $\pi$ exchange between flavored mesons, in particular charmed mesons, 
was suggested as a source of potential ``deusons" \cite{torn,ericson}. Using the deuteron binding as 
normalization, the attractive force between the $\JP = 0^-$ charmed $\D$ and its $\JP =1^-$ counterpart 
$\D^*$ was calculated for the $\D\bar {\D}^* + c.c.$ S-wave combination with total $\JPC = 1^{++}$, 
and the results compared with the enigmatic charmoniumlike state $\X(3872)$ 
\cite{pdg08,torn,cp,swanson,classics,fcthomas}. This mechanism may also play a role for the 
$\Y(4260)$ \cite{cdprl}, but expected to be suppressed in the charm-strange examples cited above. 
Thus it appears that the understanding of the dynamics of tetraquark systems is in its infancy.

As pion exchange between baryons is a well-established binding mechanism, one can expect a similar role 
to be played in meson molecule spectroscopy. The initial challenge is to clarify which exotic states 
arise due this force. Once a pattern of such states is established as a function of flavor and spin, 
it may be possible to isolate the roles of other mechanisms, such as those that play a role in the 
charm-strange and $\Y(4260)$ states. Thus the $\Z(4475)$ offers a potential theater for assessing 
various possible molecule models. Here we discuss two different charm meson molecule models of the 
$\Z(4475)$. The first, suggested by the $\D\bar {\D}^*$ molecule model of the $\X(3872)$, generalizes 
this model to include a radially-excited 2S charm meson. The second model assumes that the $\Z(4475)$ 
instead includes an orbitally excited charm meson, and is a near-threshold molecule of a $\D^*\bar {\D}_1$ pair 
\cite{mengchao, cdprl,cdt}.

\section{Molecules}

Isovector states in the charmonium sector strongly suggest molecular interpretations.
Such states may be bound states of charmed mesons -- supercharmonium, with binding driven for example 
by $\pi$-exchange forces -- or more exotic configurations of diquarks, whose binding lies in fundamental 
QCD effects. The former ``charm meson molecule" option is relatively conservative, and given that 
many molecular states are well known in the baryon sector (as atomic nuclei), this category 
appears a prime candidate for supercharmonium dynamics. Tetraquark configurations, driven by QCD, 
are less constrained and, furthermore, tend to imply the existence of many states. In contrast, 
the empirical spectrum of experimental states appears to be rather sparse.

Within the conservative supposition that pion exchange provides the attractive binding force between pairs 
of charmed mesons, there are two immediate possible origins for the $\Z(4475)$. First is traditional 
$\pi$ exchange in a pion-hadron P-wave, as occurs in pion exchange between pairs of nucleons in atomic nuclei. 
This provides potential bound states with binding energies of MeV scale, as in the case of the $1^+(3872)$ 
bound state of $\D\bar {\D}^*$. The $\Z(4475)$ as a $1^+$ state is {\it a priori} consistent with 
a radial analog of that system, with $\pi$ exchange providing binding between linear superpositions of 
$\D(\2S)\bar {\D}^*(\1S)$ and $\D(\1S)\bar {\D}^*(\2S)$ states, which we call the $\D\bar {\D}^*(1S,2S)$ 
system. The proximity of the $\X(3872)$ to $\D^0\bar {\D}^{*0}$ threshold is comparable to the $d-u$ quark 
mass difference, so that isospin symmetry violation occurs in $\X(3872)$ decays. For the radial analog system, 
isospin is in contrast expected to be a good quantum number. The important question of whether either or 
both of the $\I = 0,1$ sectors of the $\D\bar {\D}^*(1S,2S)$ system is predicted to support bound states 
immediately arises.

A second possibility is that deeper binding may be driven by $\pi$-exchange if the pion-hadron pair is 
in a relative S-wave. This can occur in the ${\D}^*\bar {\D}_1$ sector, as was discussed in references 
\cite{cdprl, cdt}. The ground state of such a system is consistent with the established $1^-$ $\Y(4260)$; 
P-wave excitations of this system are predicted to include bound states some 200 MeV higher in mass, 
as well as a $1^+$ state that is consistent with the $\Z(4475)$. We shall focus on these two possible 
$\pi$-exchange families of supercharmonia.

\subsection{P-wave Pion Exchange}

Pion emission vertices connect $\D$ to $\D^*$ (or charge conjugates). In the analysis of the $\X(3872)$ 
these were all 1S states, and after taking charge conjugation into account there were two independent 
configurations: $\D\bar{\D}^* \pm \bar{\D} {\D}^*$. Binding occurred in the $\I = 0$ $\JPC = 1^{++}$ 
channel (before $u,d$ mass effects were taken into account) \cite{cp, swanson}.

In the case in which one charmed meson is in a 1S state and its partner is a 2S state, four quantum states are 
available because $\bar{\D}(\2S){\D}^*(\1S)$, $\bar{\D}(\1S){\D}^*(\2S)$ and their charge conjugate modes 
are distinct. Qualitatively one would expect that $\pi$-exchange would give attraction in two of the diagonalized 
channels and repulsion in two. This diagonalization, combined with small energy denominators due to similar 
$\D(\1S)\D^*(\2S)$ and $\D(\2S)\D^*(\1S)$ thresholds, can lead to an amplification of binding. We find that 
given pion exchange, attractive interactions are expected in the $\JPG = 1^{++}$ channel with $\I= 0$ 
(as for the $\X(3872)$) as well as in the $\I = 1$ channel, with relative strengths of 3:1. Note that 
the neutral member of the $\I = 1$ channel has $\JPC = 1^{+-}$.

This model involves the following channel couplings (in the following we will normally suppress 1S labels):

\be
\D\bar {\D}^*(\2S) \to {\D}^*(\2S) \bar {\D},
\ee

\be
\D \bar {\D}^*(\2S) \to {\D}^* \bar {\D}(\2S),
\ee
and
\be
\D(\2S) \bar {\D}^* \to {\D}^* \bar {\D}(\2S).
\ee
These span a four-by-four effective potential with identical two-by-two subblocks. The entries in the 
potential matrix can be obtained from a nonrelativistic reduction of a one pion-exchange interaction 
that coupled point-like pions to quarks:

\be
{\cal L} = \frac{g}{\sqrt{2}f_\pi} \bar \psi \gamma^\mu \gamma_5 \vec \tau \cdot \partial_\mu \vec \pi \psi.
\ee
Here $f_\pi = 92$ MeV is the pion decay constant, $\tau/2$ is an SU(2) flavor generator, and $g$ is a 
coupling to be determined. The effective potential is derived by projecting the quark level interactions 
onto hadronic states in the nonrelativistic limit. In the case of pseudoscalar-vector states one obtains 
\cite{torn}

\be
V_{\pi} = - \gamma V_0 \left[ \pmatrix{1 & 0 \cr 0 & 1}C(r) + \pmatrix{0 & -\sqrt{2} \cr
-\sqrt{2} & 1 } T(r) \right]
\label{Vpi}
\ee
where

\be
C(r) = {\mu^2\over m_\pi^2} {{\rm e}^{-\mu r} \over m_\pi r},
\label{C}
\ee
\be
T(r) = C(r)\left( 1 + {3\over \mu r} + {3\over (\mu r)^2} \right),
\label{T}
\ee
and
\be
V_0 \equiv {m_\pi^3\over 24 \pi}{g^2 \over f_\pi^2} \approx 1.3\ {\rm MeV}.
\ee
The matrix elements refer to S- and D-wave components of the pseudoscalar-vector system, in analogy 
with the deuteron.

The strength of the interaction has been fixed by comparing to the $\pi \NN$ coupling constant through 
the relationship $g_{\pi \NN}^2/4\pi = 25/18\cdot m_\pi^2 g^2/f_\pi^2$. This allows a prediction of 
the ${\D}^*$ decay width which is in good agreement with experiment \cite{torn,swanson}. The parameter 
$\mu$ is typically the pion mass, however one can use it to incorporate recoil effects in the potential 
by setting $\mu^2 = m_\pi^2 - (m_V-m_{Ps})^2$.

Finally, the coupling $\gamma$ is a spin-flavor matrix element that takes on the following values: 
$\gamma = 3$ for $\I = 0$, $\C = +$; $\gamma = 1$ for $\I = 1$, $\C = -$; $\gamma = -1$ for $\I = 1$, 
$\C = +$; and $\gamma = -3$ for $\I = 0$, $\C = -$. The most attractive binding is in the $\JPG = 1^{++}$ 
channel. Thus the isoscalar $\JPC = 1^{++}$ channel is the most likely to form bound states, while 
the isovector negative charge conjugation channel, whose neutral member has $\JPC = 1^{+-}$, is the next 
most likely.

The tensor potential of Eq. \ref{T} is an illegal quantum mechanical operator and must be regulated, 
typically with a dipole form factor. The regulator scale, $\Lambda$ can be fixed by comparison with 
nuclear physics; for example $\NN$ interactions yield preferred values for $\Lambda$ in the range 
0.8 GeV to 1.5 GeV depending on model details. Alternatively, reproducing the deuteron binding energy 
requires $\Lambda \approx 0.8$ GeV. A value of $\Lambda = 1.2$ GeV which is appropriate for $\D$ mesons 
was used in Refs. \cite{torn,swanson} and this is taken as the canonical cutoff in the following.

Note that the coupling is fixed for interactions between all possible mesons. Differing internal 
mesonic structure is reflected in the smearing of the potentials of Eq. \ref{C} and \ref{T}. 
Thus application to pion exchange between excited state mesons can differ in form. However this is 
difficult to implement within a simple dipole regulation scheme; we therefore restrict investigation 
of this effect to varying the dipole cutoff $\Lambda$ or the effective coupling, $g$.

The channels considered in our computation are $(\D\bar {\D}^*(\2S))_S$, $(\D\bar {\D}^*(\2S))_D$, 
$(\D(\2S) \bar {\D}^*)_S$, $(\D(\2S)\bar {\D}^*)_D$, and $J/\psi \,\pi(1300)$ for the isovector state, or 
$\psi(\2S)\, \omega$ and $J/\psi\, \omega(\2S)$ for the isoscalar molecule. The mass of the $\D(\2S)$ 
state has been fixed at 2580 MeV, the $\D^*(\2S)$ mass was taken to be 2640 MeV \cite{D-models}, 
and the $\omega(\2S)$ state mass was set to 1425 MeV.

\subsection{Quark Exchange Induced Effective Interaction}

It is also of interest to examine short-range quark-exchange interactions between the relevant mesons pairs. 
These interactions can be estimated with nonrelativistic quark dynamics mediated by an instantaneous 
confining interaction and a short range spin-dependent interaction. The color structure of the quark-(anti)quark 
interaction is taken to be the usual $\lambda\cdot\lambda$ quadratic form of perturbation theory. 
This is an important assumption for multiquark dynamics, which has received support from lattice
computations for both confinement \cite{bali} and multiquark interactions \cite{bb}.
The final form of the interaction Hamiltonian is thus taken to be

\be
\label{Vij}
H_I = \sum_{i<j}{\bm{\lambda}(i) \over 2}\cdot {\bm{\lambda}(j) \over 2} \left \{
{\alpha_s \over r_{ij}} - {3\over 4} br_{ij}
- {8 \pi \alpha_s \over
3 m_i m_j } \bm{S}_i \cdot \bm{S} _j \left ( {\sigma^3 \over
\pi^{3/2} } \right ) e^{-\sigma^2 r_{ij}^2}
\right \},
\ee
where ${\bm{\lambda}}$ is a color SU(3) Gell-Mann matrix, $\alpha_s$ is the strong coupling constant, 
$b$ is the string tension, $m_i$ and $m_j$ are the interacting quark or antiquark masses, and $\sigma$ is a
range parameter in a regulated  spin-spin hyperfine interaction. The parameters were obtained from a 
global fit to the meson spectrum and are $\alpha_s = 0.59$, $b = 0.162$ GeV$^2$, $\sigma = 0.9$ GeV, 
and $0.335$, $0.55$, and $1.6$ GeV for up, strange, and charm quark masses respectively. Relevant 
meson masses obtained from this model are $\rho = 0.773$ GeV, $J/\psi = 3.076$ GeV, $\D = 1.869$ GeV, 
and $\D^* = 2.018$ GeV, in reasonably good agreement with experiment.

Meson-meson interactions are obtained by computing the Born-order scattering amplitude for a given process 
\cite{ess,BS} and equating this amplitude to an effective amplitude involving point-like mesons. Because 
of the color factors in Eq. \ref{Vij} this amplitude necessarily involves an exchange of quarks between 
the interacting mesons. Thus the leading order $\bar {\D} {\D}^*(\2S)$ interaction couples this channel 
to hidden charm channels such as $\psi(\2S) \, \omega$, $J/\psi\, \omega(\2S)$, and $J/\psi\, \pi(1300)$.

\subsection{P-wave Pion Exchange -- the ${\D}^*\bar {\D}(\1S,\2S)$ System}

For reference, the thresholds assumed for this system are $\bar {\D} {\D}^*(\2S)$ = 4509 MeV, 
$\bar {\D}(\2S) {\D}^*$ = 4590 MeV, $\rho \psi'$ = 4461 MeV, and $\omega\psi'$ = 4469 MeV. 

Given our specific choices above for the dipole cutoff, pion-quark coupling constant, and quark model parameters 
we find no isovector bound state with $\JPG = 1^{++}$. However, an isoscalar resonance with this $\JPG$ does 
exist at 4480 MeV. Since this is a radial (isoscalar) analog of the $\X(3872)$ we call it the $\X'(4480)$. 
The existence of this state is somewhat sensitive to the details of the one-pion-exchange force we assume 
in our model, but it is not sensitive to the quark-exchange interaction assumed. 

Some examples of model variation tests are as follows. If the hidden charm channels are removed from 
the system, one obtains an $\X'$ at a very slightly higher mass of 4482 MeV. Introducing an artificially 
high $\omega$ mass of 880 MeV, we again find an $\X'$ bound state at 4480 MeV. Reducing the $\omega$ mass 
to its physical value of 780 MeV leaves the $\X'$ mass at approximately 4480 MeV, but the state becomes 
a resonance since it lies above the $\omega \psi'$ threshold. Doubling the strength of the effective 
quark-exchange interactions increases the binding, and the $\X'$ mass drops to 4473 MeV.  

Variations in the pion-exchange model can be examined by changing the value of the dipole cutoff or by allowing 
the pion-quark coupling $g$ to change. Although this coupling is fixed by the model, a simple way to 
incorporate hadronic form factors at low momentum transfer is to adjust these couplings to match the 
form factors predicted by the \3P0 decay model, which are shown in Fig. 
The major effect of this modification is to rescale the $\D\bar {\D}^*(\2S)$ coupling  by about a factor of 1/4. 
In this case, an isoscalar bound state still appears, but with a mass just below threshold. 

\begin{figure}[ht]
\includegraphics[width=8cm,angle=0]{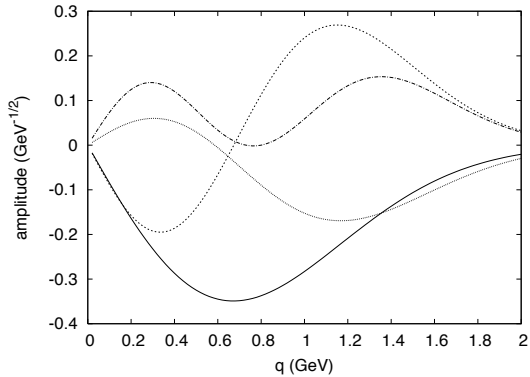}
\caption{\3P0 Model Decay Amplitudes. Solid line: ${\D}^*\to \D\pi$; dashed line: $\D(\2S)\to {\D}^*\pi$; 
dotted line: ${\D}^*(\2S)\to \D\pi$; dash-dot line: ${\D}^*(\2S)\to \D(\2S)\pi$.}
\label{fig-gap}
\end{figure}

Finally, with our standard parameter set the following component strengths are obtained for the $\X'$:
$\bar {\D}(\2S) {\D}^*|_S = 12\%$,
$\bar {\D}(\2S) {\D}^*|_D = 8\%$,
$\bar {\D} {\D}^*(\2S)|_S = 58\%$, 
$\bar {\D} {\D}^*(\2S)|_D = 14\%$, 
$\psi(\2S)\, \omega = 5\%$, and
$J/\psi \, \omega(\2S) = 2\%$. The rms radius of the state is approximately 1 fm, absent the coupling 
to the continuum.

It is of interest to examine whether an isovector state with $\Z(4475)$ quantum numbers can be generated 
within reasonable model variations. The quark exchange potential corresponding to $\D\bar {\D}^*(2S)$ or 
$\D(\2S)\bar {\D}^* \to J/\psi \pi(1300)$ is quite weak, and this interaction alone appears insufficient 
to generate a $\Z(4475)$ bound state. Increasing the dipole cutoff to 2.5 GeV would give such an isovector 
state at 4475 MeV, however we do not regard this as reasonable since it would also predict isoscalar 
partner states at very low masses of 2530 and 4080 MeV. Alternatively, increasing the effective pion-quark 
coupling by 40\% is sufficient to generate a $\Z(4475)$ isovector bound state with the required 
$\JPG = 1^{++}$ (thus giving a neutral member with $\JPC = 1^{+-}$). With these parameters we would predict 
isoscalar states with masses of 4225 and 4495 MeV. Thus, a $\Z(4475)$ state can be accommodated within 
this $\pi$ exchange model, but the pion exchange interaction required is somewhat larger than that predicted 
using our standard parameter set.

In summary, we have found that accommodating the $\Z(4475)$ as a $\D {\bar {\D}}^*({\1S},{\2S})$ 
charm meson molecule in this model implies the existence of at least one isoscalar partner, 
$\sim 100$ MeV (or more) lower in mass.

\subsection{S-wave Pion Exchange -- the ${\D}^*\bar {\D}_1$ System}

In the above examples, as in the more traditional case of the nucleon, where the $\NN\pi$ coupling is the 
source of an attractive force that helps to form the deuteron, the exchanged $\pi$ was emitted and absorbed 
in a relative P-wave with respect to each charm meson. In such cases, the binding energies that result 
are $\sim 1-10$ MeV; hence the consistency with the $\X(3872)$, which within errors is degenerate with 
the $\D^0\bar {\D}^{*0}$ threshold. Alternatively, the exchange of a pion in S-wave between pairs of 
hadrons that are themselves in a relative S-wave potentially leads to deeply bound states of these hadrons 
\cite{cdprl}. Instead of binding of a few MeV, there is now the potential for binding on the scale of 
$\sim 100$ MeV, leading to a rich spectroscopy of bound states that are far below the thresholds of 
the di-hadron channels that generate them.  We remark that both  
the $\D^*\bar{\D}_1$(broad) and the $\D^*\bar{\D'}_1$(narrow) systems will contribute to possible bound states. 
The broad $\D_1$ couples strongly to $\D^*\pi$ and thus presumably dominates the state, which implies that 
the bound state will also be broad.

The phenomenology of S-wave pion-exchange in the $\D^*\bar {\D}_1$ system was discussed at length in \cite{cdt}. 
The modelling of this dynamics is not well constrained, and so only general remarks about the spectrum 
are possible. For our purposes it is sufficient to note that with reasonable parameters the $\Y(4260)$ 
and $\Y(4360)$ $1^{--}$ states are consistent with expectations for the 1S and 2S levels of the 
$\D^*\bar {\D}_1$ charm meson molecule bound state spectrum. 

The 1P level of this system, with both $\I = 0$ and $\I = 1$ members, was found to lie near threshold, 
possibly bound at the $\sim 1$ MeV scale. Thus the $1^+$ $\Z(4475)$ state is intriguingly consistent 
with expectations for the P-wave $\D^*\bar {\D}_1$ charm molecule spectrum. Modulo spin-dependent 
mass shifts, this picture would predict companion P-wave $\D^*\bar {\D}_1$ states with $\JP = 0^+$ and $2^+$. 
The spectrum is dependent on aspects that are {\it a priori} indeterminate, such as the modelling of 
pion -- charm meson form factors, but it is intriguing that bound states can occur with either 
charge conjugation, thus both $\JPG = 1^{++}$ and $1^{+-}$ are possible.

Were a spectrum of states to be confirmed, more detailed modelling would be merited, including a study of the 
effects of form factors and of intermediate states containing on-shell pions. As discussed in \cite{cdt} and 
\cite{meissner}, the presence of on-shell pions can wash out sharp states; however, the presence of 
strength in the continuum of $\D\bar {\D}3\pi$ as well as $\D\bar{\D} 2\pi$ should be a robust signature 
of a $\D^*\bar {\D}_1$ molecule
throughout the region, in contrast to the $\bar {\D} {\D}^*(\1S,\2S)$ model, in which $\G = +$
bound states decay dominantly only to $\D\bar{\D} 2\pi$ final states.

\section{Tests}

Here we consider two broad classes of supercharmonium:  
(i) states generated by $\pi$-exchange in either $\D\bar {\D}^*$ or ${\D}_1\bar {\D}^*$ systems, as above, and 
(ii) states that arise at or near an S-wave threshold for charmed meson pairs. 
In some cases these two coincide. Thus the $1^{++}$ state $\X(3872)$ is both at $\D\bar {\D}^*$ threshold 
and is also a natural consequence of $\pi$-exchange binding. The $\Y(4260)$ however is near $\D\bar {\D}_1$ 
threshold, but can also be generated by $\pi$-exchange in the ${\D}^*\bar {\D}_1$ system. The $\Y(4360)$ 
and $\Z(4475)$ are both not far from the $\D^*\bar {\D}_1$ threshold, but are also not far from the 
$\D(\2S)\bar {\D}^*$ and $\D\bar {\D}^*(2S)$ thresholds, in which case $\pi$-exchange in this system 
is also possible. In this section we consider some more general tests that follow from these various 
possibilities, and we propose experiments that may help to distinguish among them.

\vskip 0.2in

(a) {\Y(4260)}
\vskip 0.1in

The most immediate consequence is that $\JPG = 1^{--}$ states associated with the $\D\bar {\D}_1$ or $\D^*\bar{\D}_0$ 
threshold should have strength in channels where $\D\bar{\D}$ are accompanied by an {\it even} number of pions, 
whereas if their dynamics is predominantly coupled with either $\D^*\bar {\D}_1$ or $\D(\2S)\bar {\D}^*$ 
and $\D\bar {\D}^*(\2S)$, their decays will have a preference for $\D\bar{\D}$ plus an {\it odd} number of pions. 
This should be a sharp discriminator for the $\Y(4260)$, which has so far only been clearly identified 
in $\psi \pi\pi$, with no evidence in $\D\bar{\D}, \D\bar{\D}^*$ or $\D^*\bar{\D}^*$. Any link to the 
$\D\bar {\D}_1$ threshold should generate a significant signal in $\Y(4260)$ $\to \D^*\bar{\D}\pi \to 
\D\bar{\D}\pi\pi$. This remark, which should be qualitatively true in general, has been examined in a detailed 
model in Ref. \cite{qiang14}.

To the best of our knowledge, the only data published on $\D\bar{\D}2\pi$ and $\D\bar{\D}3\pi$ relevant to the 
$\Y(4260)$ are in Refs. \cite{prl1,prl2}. (Data on $\D\bar{\D}2\pi$ from CLEO in Ref. \cite{cleo} 
is due to $\D^*\bar {\D}^*$, and advertises no signal for $\D\bar {\D}_1$). However, these papers give data 
only at a single fixed energy, $\sqrt{s} = 4.26$ GeV. What are required for our proposed tests are 
(i) a scan across $\sqrt{s}$ to see if these signals rise and fall in a manner consistent with the 
$\sim 90$ MeV width of the $\Y(4260)$ structure previously seen in $\psi \pi\pi$, and 
(ii) a determination of the ratio of strengths in the two modes $\D\bar{\D}2\pi$ and $\D\bar{\D}3\pi$. 
We urge that this sharp test be carried out.

To satisfy G-parity, the $\JP$ of the resulting $\D\bar{\D}$ and pion systems differ in these two cases.  
For $\D\bar{\D}2\pi$, the $\D\bar{\D}$ may appear in $1^{--}$ with $\pi\pi$ in $\JPC = 0^{++}$, analogous 
to the presently observed $\psi \pi\pi$ decays. In the particular case of a $\D\bar {\D}_1$ threshold state, 
there should be strong affinity for this to occur through
$\D\bar {\D}_1 \to \D\bar {\D}^*\pi \to \D\bar {\D}\pi\pi$ (as in Ref. \cite{qiang14}). For $\D\bar{\D}3\pi$, 
in contrast, the $\D\bar{\D}$ in $1^{--}$ can accompany $\eta$; alternatively, the $\D\bar{\D}$ may form 
$\JPC =0^{++}$, with the $\JPC(3\pi) = 1^{--}$. Thus a search or limit on $\Y(4260) \to \psi \eta$ in P-wave and 
for the S-wave decay mode $\D\bar{\D}\omega$ are relevant. In principle these may be able to distinguish 
the role of $\D\bar {\D}_1$ and $\D^*\bar {\D}_1$ in the dynamics of the $\Y(4260)$: a $\D\bar {\D}_1$ 
threshold effect is dominantly the former; a deeply bound $\D^*\bar {\D}_1$ will have significant preference 
for the latter; a hybrid vector will couple to both of these, with a preference for the latter, 
but branching ratios are model dependent due to the $\D\bar{\D}3\pi$ being driven by virtual 
intermediate states. Nonetheless, a richer data set, as advocated here, may be able to eliminate 
one or more scenarios. 

Superficially the data from Refs. \cite{prl1} and \cite{prl2} suggest that the strengths of 
$\D\bar {\D}^*\pi$ and $\D^*\bar {\D}^*\pi$ are not dissimilar. Although the published experiments 
do not enable immediate extraction of the actual ratio of $\sigma(\D\bar {\D}^*\pi)/\sigma(\D^*\bar {\D}^*\pi)$ 
at 4.260 GeV, this ratio is approximately 2.9 \cite{CZY}.  Nevertheless, one cannot infer the ratio 
of matrix elements for $\Y(4260)$ coupling to $\D\bar {\D}_1$ to $\D^*\bar {\D}_1$ because
(i) the data do not show what fraction of $\D^*\pi$ is actually ${\D}_1$; 
(ii) phase space effects and the dynamics of couplings to states that are below threshold are nontrivial.
Thus, measuring the ratio of $\D \bar {\D}_1$/$\D^*\bar {\D}_1$ over a range of $\sqrt{s}$ would be useful 
as it would aid in disentangling these effects.

\vskip 0.2in

(b) \Z(4475)
\vskip 0.1in

The $\Z(4475)$ state has $\I = 1$ and decays to $\psi' \pi$, and is therefore presumably $\JPG = 1^{++}$ 
(its neutral member thus have $\C = -$). While both $\D\bar {\D}^*(\1S,\2S)$ and $\D^*\bar {\D}_1$ 
bound state possibilities admit the presence of $\JPG = 1^{++}$, and can accommodate the $\Z(4475)$, 
$\D^*\bar {\D}_1$ molecules allow also a $\JPG = 1^{+-}$ state in this mass region. The $\D^*\bar {\D}_1$ 
model also predicts $0^+$ and $2^+$ isovector and isoscalar bound states, of both G-parities. Clearly, 
searching for these states has high priority. Confirmation or elimination of this rich spectroscopy 
is perhaps the sharpest test for the role of this $\D^*\bar {\D}_1$ dynamics.

For the particular case of the $\JP = 1^+$ states, we summarize the situation in Table \ref{tab:iso}.

The $\D\bar {\D} \pi\pi\pi$ channel should be important for the $\Z(4475)$ state, whether it is associated 
with the $\D\bar {\D}^*(\1S,\2S))$ or the $\D^*\bar {\D}_1$ systems. Both of these give attraction 
in the $\G = +$ sector, thus $\JPC = 1^{++}; \I = 0 $ (e.g., $\X(3872)$ and its radial analog) and 
$\JPC = 1^{+-}; \I = 1$, such as $\Z(4475)$. 
The structure of the $\D\bar{\D}\pi\pi\pi$ decay mode can be used to distinguish $\D^*\bar{\D}_1$ and $\D\bar{\D}^*(\1S,\2S)$ systems because the former will display $\D^*\bar{\D}^*$ resonant structure while the latter will have a complex resonant structure dominated by $\D\bar{\D}^*$. Both cases yield $\Z(4475)$ states that are broad, due to the decay of broad $\D_1$ or $\D_0$ virtual states. This is commensurate with the measured width of $172(37)$ MeV\cite{4475}.
For the $1^{++}$ case, the $3\pi$ system is dominated by the $\omega$, 
whereas for the $1^{+-}$, they it form a broad continuum such as $\pi^*(0^{-+})$ or $\rho\pi$ 
(as in $a_1$).

Another consequence of both the $\D\bar {\D}^*(\1S,\2S)$ and $\D^*\bar {\D}_1$ interpretations of the 
$\Z(4475)$ as an $\I = 1$, $\JPG = 1^{+-}$ state is that $\pi$-exchange necessarily binds an $\I = 0$ 
partner at least some tens of MeV lower in mass. Finding this state is crucial for these molecular 
hypotheses. In both cases, the deeper bound $\I = 0$ $1^{++}$ would make E1 radiative transitions 
to $\gamma \psi'$ and to $\gamma J/\psi$, so we recommend searches in the invariant mass distributions 
of those final states, as well as searches in $J/\psi \omega$, which can be manifest in $J/\psi \pi \gamma$. 

\begin{table}[ht]
\begin{tabular}{l|l|l}
\hline\hline
    & $G=+$   & $G =-$ \\
\hline
  $I=0$  & $1^{++}$ \ $J/\psi\, 3\pi (\omega)$\phantom{$\pi$} \ S  & $1^{+-}$ \ $J/\psi\, 3\pi (\eta)$ \ S\\
         &                                          & $1^{+-}$ \ $J/\psi\, 2\pi$\phantom{($\eta$)} \ P \\
\hline
  $I=1$  & $1^{+-}$ \ $J/\psi\, \pi$ \phantom{$3(\rho\pi)$} \ S   &  $1^{++}$\, $J/\psi$\, $2\pi (\rho)$ \hspace{-1pt} P  \\
         & $1^{+-}$ \ $J/\psi\, 3\pi (\rho \pi)$ \hspace{0.4pt} \ P  &   \\ 
\hline\hline
\end{tabular}
\caption{Isospin and G-parity eigenstates with $\JP = 1^+$, and some decay channels with $J/\psi$. 
The $\JPC$ of neutral members is also shown, together with the partial wave (S or P) of the decays.}
\label{tab:iso}
\end{table}

Radiative transitions between different levels of the $\D^*\bar {\D}_1$ bound states are possible, 
and can be used to gain information. For example, in the $\D^*\bar {\D}_1$ scenario, a $\JPC = 1^{++}$ 
charge conjugate partner of the $\Z(4475)$ can decay to the $\Y(4260)$ via the process \cite{cdt} 
$[{\D}_1\bar {\D}^*]_P \to [{\D}_1\bar {\D}^*]_S + \gamma$. Thus the charged partner state could 
be revealed in the $J/\psi \pi\pi \gamma$ channel. A similar radiative transition can also occur 
in the $\D\bar {\D}^*(\1S,\2S)$ system for the isoscalar $\X'$, as $\D\bar {\D}^*(\2S) \to \D\bar {\D}_1 \gamma$, 
where the $\Y(4260)$ is interpreted as a $\D\bar {\D}_1$ state. 

Prospects for studying these channels do not appear promising at BESIII because the highest center of mass 
energy available is 4.6 GeV. Furthermore, data at this energy are sparse \cite{thankryanmitchel}. 
The situation at Belle is likely to be worse. However, $\D\bar {\D}\pi\pi$ and $\D\bar {\D}\pi\pi\pi$ data 
exist, as in \cite{prl1,prl2}, and should be extended over as wide a range of $\sqrt{s}$ as possible.  
Similarly, there are plans to study $\D\bar {\D} n\pi$ with charged pions at LHCb \cite{thankmikewilliams}.

\section{Conclusions} 

The interplay of long-range pion exchange forces and charm meson pair dynamics is likely to play a role 
in many supercharmonium states. It is possible, for example, that the $\Z(4475)$ is either a 
${\D}^*\bar {\D}_1$ state dominated by long-range pion exchange, or a $\D\bar {\D}^*(\1S,\2S)$ state 
with important short-range components. In the first case one expects a nearby isoscalar molecular state, 
and a rich spectrum of $\JP = 0^+$ and $2^+$ states with both positive and negative G parity. 
In the latter case an isoscalar state with $\JPC = 1^{++}$ well below the $\Z(4475)$ is expected.

If additional information warrants it, the model considerations presented here could be improved by considering 
form factors associated with the pion coupling to various constituent mesons, the possibility that 
pions go on-shell, and the presence of left hand cuts \cite{meissner}. However, at present, general 
strategies appear to be more useful than detailed analyses, which would introduce further 
model-dependent uncertainties.

We have suggested that searching for the $\Z(4475)$ and related states in $\psi(2S)n\pi$, $\D\bar{\D}$$n\pi$, 
and $J/\psi n\pi$ modes can provide crucial discriminating information on the nature of these states. 
The nature of the $\Y(4260)$ may be clarified when data on $\D\bar{\D} 2\pi$ (excluding $\D^*\bar {\D}^*$) 
and $\D\bar{\D} 3\pi$ are established through the 4.26 GeV region, and the relative importance of (virtual) 
${\D}_1\bar {\D}$ and ${\D}_1\bar {\D}^*$ states is established. Furthermore, whether the $\D\bar{\D}\pi\pi\pi$ system is dominated by $\D^*\bar{\D}^*\pi$ or $\D \bar{\D}^*\pi\pi$ can distinguish $\D{\bar D}_1$ and $\D\bar{\D}*(\1S,\2S)$ systems. Data at BESIII implicitly contain 
this information, but only at one energy; extending these data across a range of energies and extracting 
the $\D_1$ signal would be most useful.

The way in which the dynamics suggested here is realized in other flavor systems is of interest.  
Certainly, one expects greater binding in the $b$-quark analogs of all the systems considered here: 
$\B\bar {\B}^*$, ${\B}^*\bar {\B}_1$, and $\B\bar {\B}^*(\1S,\2S)$. In the case of $s$-quark analog states, 
the pion exchange model is well justified. However, the higher kinetic energies of strange states relative 
to charm states changes the computation significantly, so that $\K\bar {\K}^*$ states do not bind with pion 
exchange only \cite{torn}. Furthermore, the near degeneracies within the $\D\bar {\D}^*(\1S,\2S)$ system 
that give small energy denominators and enhance binding are not present in the strange sector; the thresholds 
$\K\bar {\K}^*(\2S)$ (at 1905 MeV) and $\K(\2S)\bar {\K}^*$ (at 2317 MeV) 
are far from degenerate.

\acknowledgments
We are grateful to Qiang Zhao for discussions of BESIII results. This research was supported in part by the 
U.S. Department of Energy under contract DE-FG02-00ER41135 (Swanson).

\end{document}